\documentclass[fleqn,usenatbib]{mnras}

\usepackage{newtxtext,newtxmath}

\usepackage[T1]{fontenc}
\usepackage{ae,aecompl}
\usepackage[framemethod=tikz]{mdframed}

\usepackage{graphicx}	
\usepackage{amsmath}	
\usepackage{amssymb}	
\usepackage{tabularx}

 \newenvironment{conditions*}
  {\par\vspace{\abovedisplayskip}\noindent
   \tabularx{\columnwidth}{>{$}l<{$} @{\ : } >{\raggedright\arraybackslash}X}}
  {\endtabularx\par\vspace{\belowdisplayskip}}

\title[Gain Stabilization for Radio Receivers]{Gain Stabilization for Radio Intensity Mapping using a Continuous-Wave Reference Signal}

\author[A. Pollak et al.]{
Alexander W. Pollak,$^{1}$\thanks{E-mail: alexander.pollak@physics.ox.ac.uk}
Christian M. Holler,$^{2}$
Michael E. Jones$^{1}$ and Angela C. Taylor$^{1}$
\\
$^{1}$Sub-department of Astrophysics, University of Oxford, Denys Wilkinson Building, Keble Road, Oxford OX1 3RH, UK\\
$^{2}$Munich University of Applied Sciences, Lothstra\ss e 34, Munich 80335, Germany\\
}

\date{Accepted XXX. Received YYY; in original form ZZZ}

\pubyear{2019}

\begin{document}
\label{firstpage}
\pagerange{\pageref{firstpage}--\pageref{lastpage}}
\maketitle

\begin{abstract}
Stabilizing the gain of a radio astronomy receiver is of great importance for sensitive radio intensity mapping. In this paper we discuss a stabilization method using a continuous-wave reference signal injected into the signal chain and tracked in a single channel of the spectrometer to correct for the gain variations of the receiver. This method depends on the fact that gain fluctuations of the receiver are strongly correlated across the frequency band, which we can show is the case for our experimental setup. This method is especially suited for receivers with a digital back-end with high spectral resolution and moderate dynamic range. The sensitivity of the receiver is unaltered except for one lost frequency channel. We present experimental results using a new $4$--$8.5$\;GHz receiver with a digital back-end that shows substantial reduction of the $1/f$ noise and the $1/f$ knee frequency. 
\end{abstract}




\section{Introduction}

Radio astronomy observations depend on stable receiving hardware, especially total intensity observations using a single antenna. The signal levels to be detected are often many orders of magnitude lower than the instantaneous noise power level in the system. Successful observations therefore rely on the ability to integrate down the noise over long timescales. Thermal noise with a white (flat) power spectrum will integrate down as the square root of the observing time. However, many systems exhibit long-period, correlated fluctuations, corresponding to excess noise power at low frequencies in the time-ordered data, which if uncorrected would dominate the noise level in long integrations. This correlated noise is typically due to fluctuations in the gain of the receiving system. This low-frequency noise can be dealt with either by introducing modulation schemes that push the signal of interest up in frequency away from the low-frequency fluctuations, or by actively stabilizing or cancelling out the fluctuations, or by a combination of these techniques.  For mapping relatively small areas of sky, it is often possible to scan the beam rapidly enough across the field of view to ensure that all signals of interest are in the white-noise dominated part of the data power spectrum. However, there is increasing interest in mapping very large areas of sky, for which this is not a feasible solution, as reasonable scan speeds will take longer to traverse the whole sky than the stability timescale of the receiver.

For example, it has been proposed to use the Square Kilometre Array in total power rather than interferometric mode to make a large-area, low-resolution survey of the redshifted neutral hydrogen line, integrating the signals from many galaxies within the beam area in order to trace the evolution of large-scale structure \citep{2015aska.confE..19S}. Even with slew speeds of several degrees per second it will take of order 100~s for the antennas to scan through 360 degrees of azimuth, requiring receiver stability over this timescale if the largest angular scale structures in the sky are to be recovered. For a spectral survey, it is possible to partly compensate for receiver instability by subtracting the varying continuum power. However, for large-scale continuum surveys receiver stabilization is essential. 

Recent examples of large-scale radiometric surveys include the Planck Low-Frequency Instrument (LFI) \citep{2002A&A...391.1185S} and WMAP \citep{2003ApJS..145..413J} from space, and C-BASS \citep{projectpaper} on the ground. Planck LFI, WMAP and C-BASS all use different forms of the continuous comparison receiver \citep{1959AnAp...22..140B}. This architecture, described in more detail in Section \ref{sec:stabilization_methods} below, provides good immunity from correlated noise in the amplifier chain, but at the expense of considerable increase in both complexity (two complete receiver chains per polarization) and system noise (due to extra components before the first amplifier, and noise of the reference signal). For similar surveys in the future, particularly ones involving large numbers of feeds, it would be desirable to both simplify the receiver architecture and achieve the best possible system temperature. In this paper we describe a prototype receiver design that achieves essentially the full theoretical sensitivity of a simple radiometer and good correction of correlated gain fluctuations, using a simple RF chain design and a continuous-wave (CW) calibration signal.

The paper is organized as follows. In Section \ref{sec:gain_fluctuations} we discuss the properties and characterization of receiver noise, and briefly describe the common receiver architectures used to deal with correlated noise. In Section \ref{sec:method} we describe our experimental setup and present results on the degree of correlation of gain fluctuation as a function of RF frequency, and the improvements achieved in the noise power spectrum by means of CW calibration. We present our conclusions in Section \ref{sec:conclusion}.

\section{Gain Fluctuations in Receiver Systems}
\label{sec:gain_fluctuations} 

\subsection{Receiver Sensitivity}

Receiver systems introduce both uncorrelated (white) and correlated (pink or red) noise on a detected signal. To clarify the nomenclature we use for different noise spectra, we consider the temperature noise power spectrum to have a power-law slope with  $P(f) \propto f^{-\alpha}$. White noise has a flat frequency spectrum ($\alpha = 0$), with equal power per linear bandwidth interval. White noise is caused by the thermal motion of electrons in electrical conductors and is characterized by the system temperature $T_{\rm{sys}}$ of the receiver. Pink noise, also known as $1/f$ noise or flicker noise ($\alpha = 1$), has a spectral power density proportional to $1/f$, with equal power per logarithmic bandwidth interval. Pink noise is found to represent noise properties in a wide range of physical situations, from electronic devices to biological processes, e.g. \citet{1978ComAp...7..103P}. Red (or Brown, or Brownian) noise has an even steeper spectrum ($\alpha=2$), with the low-frequency power dominating. The correlated noise that we will be referring to throughout this paper typically has a spectral slope $\alpha \sim 1$--$1.5$, which we will refer to as pink noise.  

We will regard the correlated noise in a radio receiver as being due to gain fluctuations $\Delta G$ with a pink noise spectrum, in addition to the white additive noise due to the thermal fluctuations. The practical sensitivity of a receiver is limited by both kinds of uncertainties,  and can be represented by a modified version of the radiometer equation:

\begin{equation} \label{eq:1}
    \Delta T = T_{\rm{sys}}  \sqrt{\frac{1}{\Delta \nu \cdot \tau}+\left( \frac{\Delta G(f)}{G} \right)^2}
\end{equation}
where
\begin{conditions*}
\Delta T & receiver sensitivity,\\
T_{\rm{sys}} & system temperature of the receiver,\\
\Delta \nu & receiver bandwidth, \\
\tau  & integration time,\\
\Delta G(f) & gain variations,\\
G & total mean gain of the receiving system.
\end{conditions*}

\noindent We distinguish between $\nu$ referring to the frequency or bandwidth of the RF system, and $f$ referring to the frequency domain of the post-detection time-ordered data. The variations of the system gain $\Delta G(f)$ give rise to the correlated component of the receiver noise $T_{\rm sys}\cdot (\Delta G(f)/G)$.

\begin{figure}
 \includegraphics[width=\columnwidth]{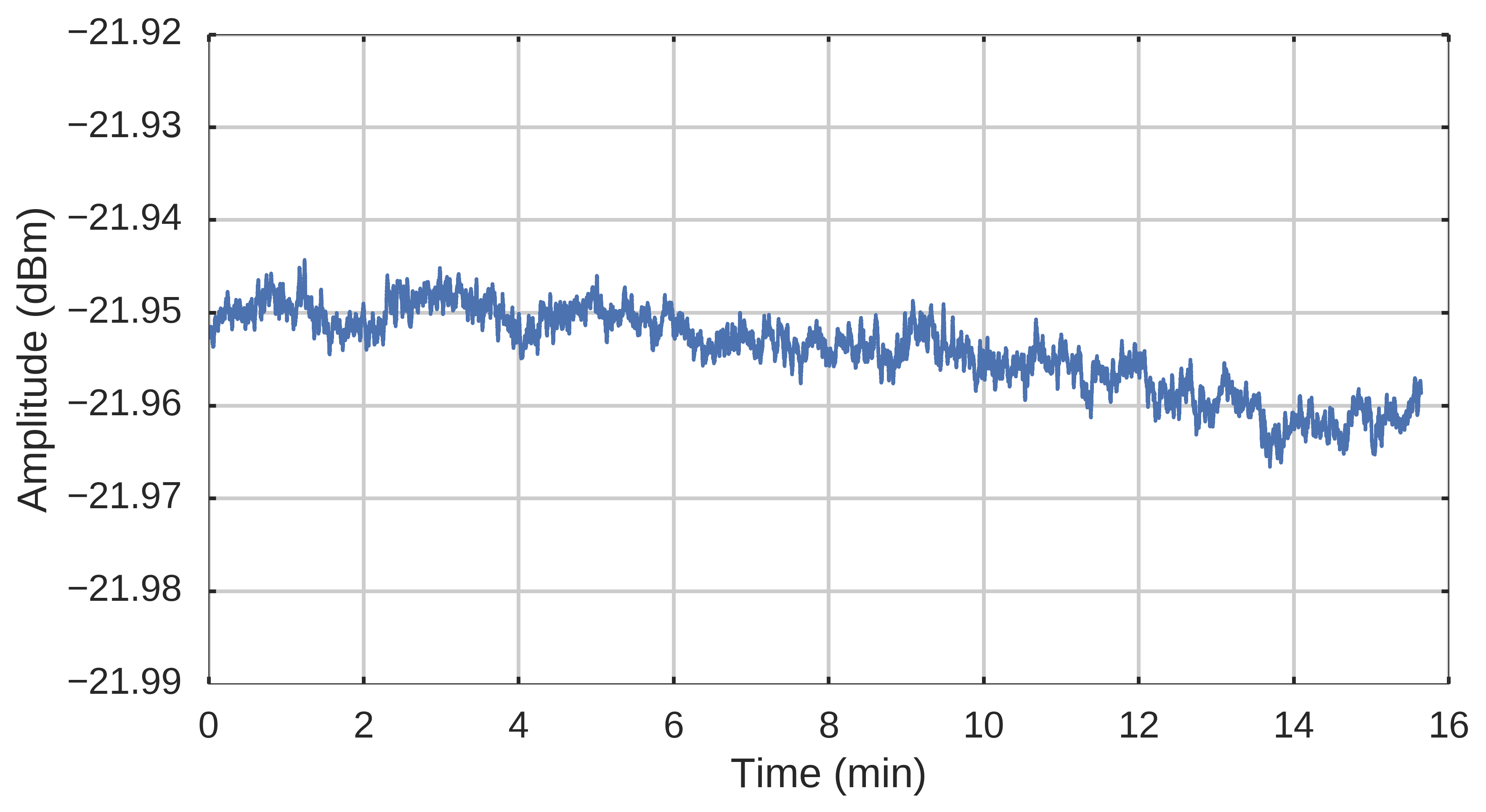}
 \caption{Measured receiver output power for $15$\;min. The signal has been low-pass filtered with a cut-off frequency of approximately $2$\;Hz in order to emphasise the gain fluctuations. The trace here is therefore dominated by pink noise.}
 \label{fig:fig1}
\end{figure}

\begin{figure}
 \includegraphics[width=\columnwidth]{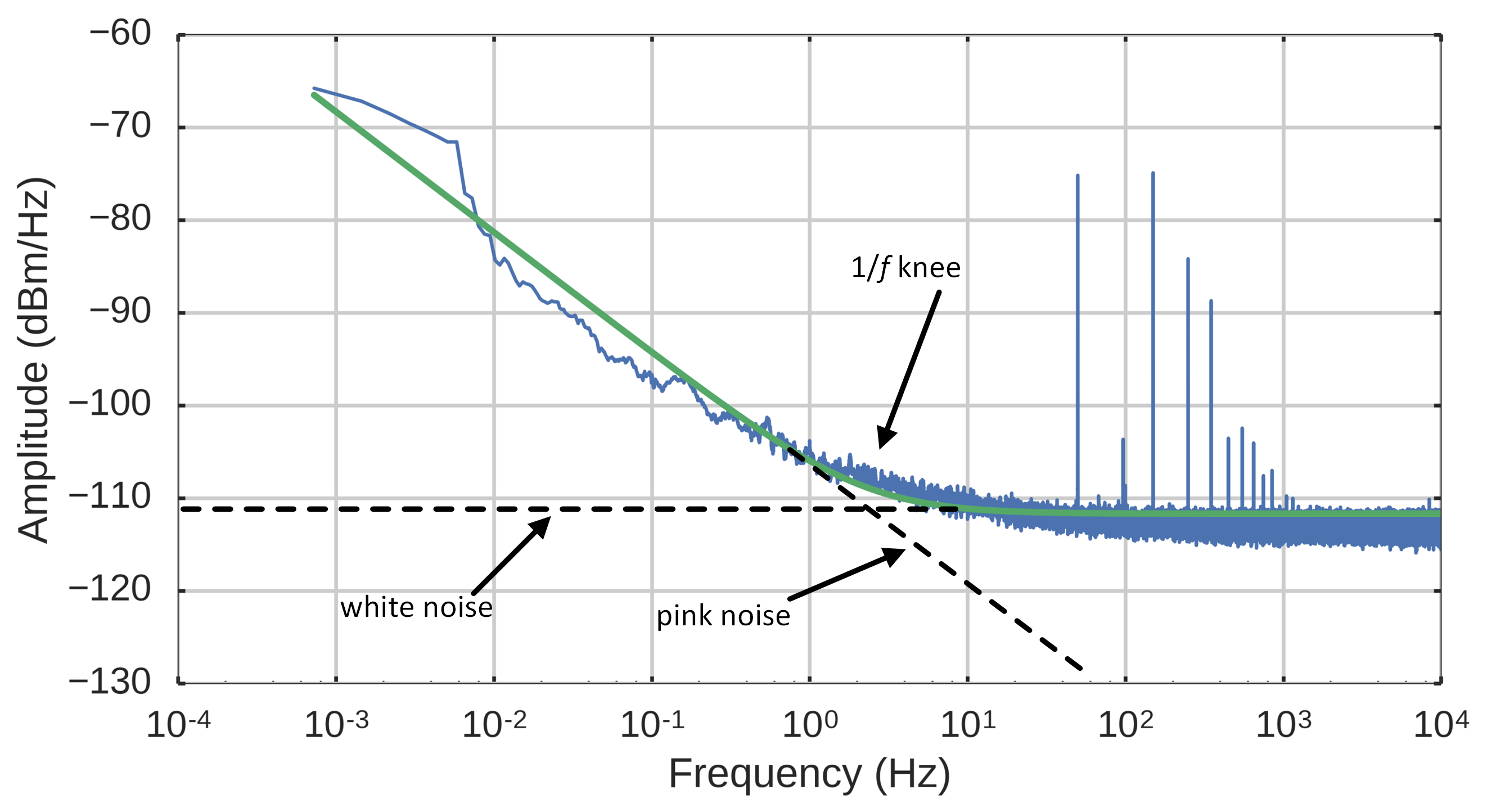}
 \caption{Noise power spectrum of the non-filtered output signal presented in Figure \ref{fig:fig1}. The spectra have been logarithmically smoothed using a moving average filter. Note the additional narrow-band signals at 50\;Hz and its harmonics due to mains-frequency pickup.}
 \label{fig:setup2}
\end{figure}

\subsection{Noise Model}

Figures \ref{fig:fig1} and \ref{fig:setup2} show an example of data from a real test receiver, showing the time ordered data and its power spectrum, respectively. The two contributions to the total noise power, white and $1/f$ noise are marked in the figure. It can be seen that for this particular receiver without any noise stabilization the integration time would be restricted to of order $1$\;s before the pink noise becomes dominant. 

We use a noise model in order to characterize the measured receiver noise power spectrum \citep{jew2017}. Power spectra such as this are often characterized by the knee frequency, at which the power contributions of the white and pink components are equal. This concept is useful, but can be inconvenient for fitting spectra as it depends on the level of both the independent noise components. For example if the white noise level goes down (an improvement in system performance) the knee frequency goes up (an apparent degradation in performance).  We therefore model the noise power density as independent white and pink components:
\begin{equation}
P= \sigma^{2}_{w} + \sigma^{2}_{c}\left(\frac{f}{f_{0}}\right)^{-\alpha},
\label{eq:NoiseModel}
\end{equation}
where
\begin{conditions*}
\sigma_{w}^{2} &  white noise level,\\
 \sigma_{c}^{2}  & correlated noise level at $f_{0}$,\\
f_{0} & a reference frequency, here 1\,Hz, \\
\alpha & power law index.
\end{conditions*}

\noindent
The green line in Figure \ref{fig:setup2} represents the results of fitting  this model to the measured power spectrum. The parameters $ \sigma_{c}^{2}$ and $\alpha$ represent the non-white noise characteristic in the power spectrum. The results of the fitting algorithm for Figure \ref{fig:setup2} are shown in Table \ref{tab:fitted_noise_model_intrinsic}.

\begin{table}
\centering
\caption{Fitted noise model parameters for the spectrum in Figure \ref{fig:setup2}.}
\label{tab:fitted_noise_model_intrinsic}
\begin{tabular}{@{}lll@{}}
\hline                                        
\multicolumn{2}{l}{Parameter}   & \multicolumn{1}{l}{Value}  \\ \hline
\multicolumn{1}{l}{$\alpha$}  &\multicolumn{1}{l|}{}  & \multicolumn{1}{l}{1.3}              \\
\multicolumn{1}{l}{$\sigma_{w}^2$}  &\multicolumn{1}{l|}{($\rm{dBm/Hz}$)}  & \multicolumn{1}{l}{$-111.7$}          \\
\multicolumn{1}{l}{$\sigma_{c}^2$}  &\multicolumn{1}{l|}{($\rm{dBm/Hz}$)}  & \multicolumn{1}{l}{$-107.3$}      \\
\end{tabular}
\end{table}

The effect of gain stabilization will typically be to reduce the amplitude of the pink noise component rather than to change its slope. We will therefore consider that the parameter characterizing the quality of correlated noise stabilization is the power level of the correlated noise $\sigma_{c}^{2}$ in comparison to the white noise power level $\sigma_{w}^{2}$. 

\subsection{Gain Stabilization Methods}\label{sec:stabilization_methods}

Several methods for receiver gain stabilization are in common practice, of which we briefly describe three in this introduction. 

A \textit{Dicke-switched}  receiver (\citet{dicke1946}, \citet{wait1967}) is based on a differential compensation approach that subtracts a known reference signal from the antenna signal. The receiver measures the sky signal for $50$ per cent of the total time and then a known stabilized reference signal for comparison for the other $50$ per cent of the time and takes the difference between the two measurements. When switching fast enough (typically $10$ to $1000$\;Hz) any gain fluctuations on a time scale larger than the switching frequency will be cancelled out. 
This increase in receiver stability however, results in a factor of two increase in the white noise level, due to fact that only half of the observing time is spent on the sky signal, and that the system temperature is increased by differencing two noise-like signals. In addition, if the sky and load powers are not perfectly matched, there will be some residual uncorrected gain fluctuations. The sensitivity for a Dicke-switched receiver is thus given by
\begin{equation}
\Delta T = T_{\rm{sys}}  \sqrt{\frac{4}{\Delta \nu \cdot \tau}+\left( \frac{\Delta G_{\rm{res}}}{G} \right)^2} \;,
\end{equation}
where $\Delta G_{\rm{res}}$ represents any residual gain fluctuations.

A \textit{pseudo-correlation} or \textit{continuous-comparison} receiver \citep[e.g.][]{mennella2003advanced, king2013c} partly overcomes this drawback by continuously comparing the sky signal to a stabilized reference source, or to an independent sky signal in a differencing radiometer such as was used in WMAP.  A hybrid is used to form the sum and difference of the sky and reference signals before amplification. Following the entire gain chain, the sky and reference signals are separated and their powers differenced.  This method allows for continuous observation but the sensitivity decreases nevertheless by a factor of $\sqrt{2}$ due to the noise on the reference signal. In addition to the need for two hybrids, twice the number of gain components are required due to the second (reference) signal chain, and twice the bandwidth has to be processed. The presence of an additional component before the low-noise amplifier also increases the system temperature. Here the sensitivity is given by
\begin{equation}
\Delta T =  T_{\rm{sys}}  \sqrt{\frac{2}{\Delta \nu \cdot \tau}+\left( \frac{\Delta G_{\rm{res}}}{G} \right)^2} . \;
\end{equation}

\textit{White noise} or \textit{noise-adding} stabilization is another method, which is particularly suitable for receivers with digital back-ends and relatively high dynamic range, and is related to the Dicke-switched system. It uses a bright broadband reference signal, usually produced by a stabilized noise diode, to track gain fluctuations. The switched reference signal is added to the antenna signal by a coupler before the first amplifier. The added reference signal allows the back-end system to determine the instantaneous gain of the receiver. This is done by constantly comparing the measured power for both signal states, with (ON) and without (OFF) the reference signal present.

In order to calculate the sensitivity of the receiver system after stabilization one has to quantify the uncertainty that is introduced by measuring the gain. This is done by repeatedly taking the difference in receiver temperature of both signal states, $T_{\rm{on}}-T_{\rm{off}}$, which results in a measurement of the instantaneous receiver gain. The uncertainties of the Gaussian noise power measurements of both states are added in quadrature, which results in a gain uncertainty for each measurement of
\begin{equation}
\frac{\Delta G}{G} = \frac{\sqrt{\left(\frac{T_{\rm{on}}}{\sqrt{\Delta \nu \cdot \tau_{\rm{on}}}} \right)^2+\left(\frac{T_{\rm{off}}}{\sqrt{\Delta \nu \cdot \tau_{\rm{off}}}} \right)^2}}{T_{\rm{on}}-T_{\rm{off}}}\;,
\end{equation}
where $\tau_{\rm{on}}$ and $\tau_{\rm{off}}$ are the durations of the ON and OFF signal states. 

Assuming a $50$ per cent duty cycle and therefore $\tau_{\rm{on}} = \tau_{\rm{off}} = \tau/2$, this simplifies to 
\begin{equation}
\frac{\Delta G}{G} = \frac{1}{\sqrt{\Delta \nu \cdot \frac{\tau}{2}}} \cdot \frac{\sqrt{T_{\rm{on}}^2 + T_{\rm{off}}^2}}{T_{\rm{on}}-T_{\rm{off}}}=
\begin{cases}
    \sqrt{\frac{2}{\Delta \nu \cdot \tau}}       & \, \text{if } T_{\rm{on}}\gg T_{\rm{off}}\\
    \sqrt{\frac{10}{\Delta \nu \cdot \tau}}       & \, \text{if } T_{\rm{on}}=2\cdot T_{\rm{off}}\\
    \infty       & \, \text{if } T_{\rm{on}}=T_{\rm{off}}\;.\\

\end{cases}
\end{equation}

\noindent This function is minimised for $T_{\rm{on}}\gg T_{\rm{off}}$. When $T_{\rm{on}}=T_{\rm{off}}$, no reference signal is present and the gain cannot be measured.

To calculate the sensitivity of this stabilized system in the case of a strong reference signal (i.e. $T_{\rm{on}}\gg T_{\rm{off}}$), $\frac{\Delta G}{G}$ in Equation \ref{eq:1} is replaced with $\sqrt{\frac{2}{\Delta \nu \cdot \tau}}$. Further, only $50$ per cent of the time is spent measuring the sky signal, since the data with the noise source switched on have very much lower statistical weight than the data with the noise source switched off. Hence the time constant $\tau$ in the first argument of Equation \ref{eq:1} is replaced by $\tau/2$. This results in a sensitivity which is identical to a Dicke-switched system with residual gain fluctuations $\Delta G_{\rm{res}}$,   
\begin{equation}
\Delta T =  T_{\rm{sys}} \sqrt{\frac{4}{\Delta \nu \cdot \tau}+\left( \frac{\Delta G_{\rm{res}}}{G} \right)^2}\;.
\end{equation}

\noindent This result is valid in the case of a duty cycle of $50$ per cent and for large reference signals $T_{\rm{on}}\gg T_{\rm{off}}$. In any other case the sensitivity of the system is further reduced (see e.g. \citet{10.2307/2098930}, \citet{wait1967}, and \citet{pollak2018} for a detailed analysis of receiver sensitivity for varying duty cycles and for different types of modulation scheme). An advantage of the white noise stabilization compared with a Dicke-switched system is however that instead of a switch, only a noise injection coupler needs to be installed in the signal chain before the first amplifier. This is likely to have less impact on the overall receiver temperature, and also allows the receiver to be used in other modes, such as as an interferometer element. 

\begin{figure*}
 \includegraphics[width=2.1\columnwidth]{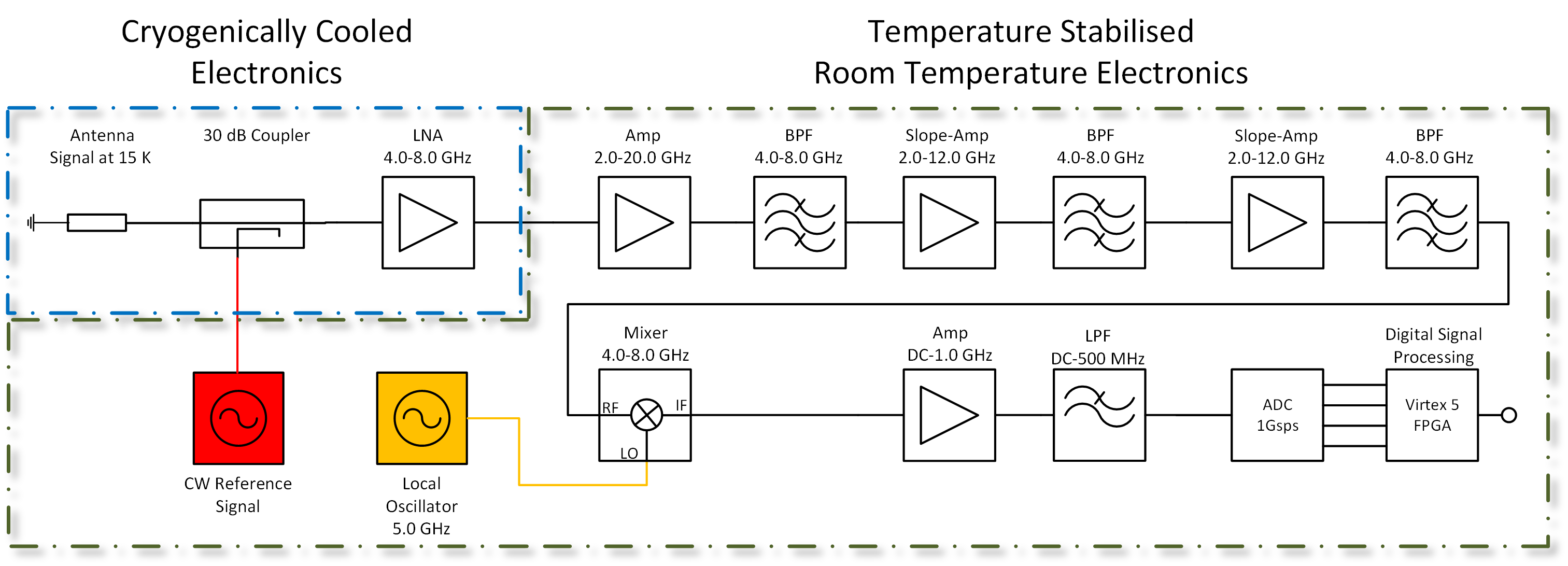}
 \caption{Schematic of the laboratory setup used to investigate the receiver gain stabilization.}
 \label{fig:fig3}
\end{figure*}

A further method of receiver gain stabilization, which is suitable for digital back-end receivers with high spectral resolution, uses a continuous-wave reference signal, which is coupled to the antenna signal before the first amplifier. This is tracked in a single frequency channel of the spectrum and used to correct for gain fluctuations in the receiving system \citep{colvin1961}. Provided that the gain fluctuations are strongly correlated across the full frequency band it is sufficient to track these fluctuations in a single channel. The resulting decrease in receiver sensitivity compared to an ideal radiometer due to this method is thus small, since most of the bandwidth contains no reference signal at all. In addition, implementation of this scheme does not require a complex receiver architecture, e.g. a complete second signal chain as is required in the continuous-comparison receiver. However, this method does require an accurately stabilized continuous wave reference source. 

In the following section we describe in detail this method of continuous-wave gain stabilization and present results from laboratory measurements showing its effectiveness.

\section{Receiver Stabilization using a Continuous-Wave Reference Signal}\label{sec:method}

Our motivation for investigating receiver stabilization methods is a project to convert a $30$-m class antenna located at the Goonhilly Earth Station in south-west England \citep{heywood2011expanding}. The antenna was formerly used for telecommunications and is being converted into an instrument suitable for radio astronomical observations in the C-band, and deep space communications at $8.4$\;GHz. For this a  cryogenically-cooled receiver system and a digital back-end have been designed \citep{pollak2018}. The total bandwidth will cover the range $4$--$8.5$\;GHz, split into sub-bands of $500$\;MHz bandwidth each, each of which is sampled by an ADC clocked at 1 GHz. The full receiver uses a multi-channel data acquisition system developed for the SKA-Low telescope \citep{TPM}, but for these tests we used a single iADC card attached to a ROACH FPGA board \citep{casper}, providing a single channel of digitization at 1 GHz clock rate and 8 bits depth.

\subsection{Method overview}
\label{sec:method_overview}

The method we present here to track and correct for gain fluctuations in a receiver system uses a continuous wave (CW) reference signal within the spectral band of the receiver. Similar methods have been presented early in the development of radio astronomy receivers, e.g. the \textit{Pilot Signal Receiver} of \citet{colvin1961}, in which the CW signal is placed outside the passband of interest. However, with the advent of digital back-ends with a high spectral resolution the advantages of this method can be fully utilized. The signal has to be injected as early into the signal path as feasible, ideally right after the feed horn, since it is only possible to correct for gain fluctuation that appear after the injection point. This method uses the assumption that the gain fluctuations are not frequency dependent across the band of the receiver. We will show that this assumption applies for the receiver under consideration. Note that for a heterodyne receiver that splits the full band into several sub bands, a CW signal needs to be injected for each sub band that has independent gain components. This can be done using either multiple fixed-frequency sources or a frequency-agile source.

The gain of the receiver is determined by repeatedly measuring the power level of the CW signal at a frequency faster than the $1/f$ frequency knee of the noise power spectrum. The advantage of this method is that the reference signal is band-limited to one channel of the spectrometer. Even a relatively low-power CW signal will have a high spectral power density and hence yield a high signal-to-noise level in the corresponding narrow frequency channel. The input power level of the analogue-to-digital converter (ADC) is therefore almost unaltered and a high dynamic range input to the ADC is not necessary. 

This method allows for continuous observation without the need of a broadband reference signal and therefore the sensitivity of the receiver is only slightly reduced due to the loss of one single frequency channel. However, high channel isolation in the implementation of the digital spectrometer is necessary. In our case this is realised by a polyphase filter bank which implements a finite impulse response filter (see e.g. \citet{price2016}, \citet{schafer1973design}, and \citet{crochiere1983multirate} for a more detailed analysis of the implementation and functionality of a polyphase filter bank and finite impulse response filter). The isolation between adjacent channels is above $60$\;dB, and for more widely spaced channels above $80$\;dB.

The accuracy of this method is limited by either the thermal noise
level on the gain measurement, or on the stability of the CW signal
used as a reference, whichever is the larger. We consider first the
thermal noise. The signal-to-noise of the gain measurement is given by
the ratio of the noise power in a single channel to the injected
signal power. The noise on a power measurement in a single channel of width $\Delta \nu_{\rm ch}$ in an integration time $\tau$ is

\begin{equation}
\Delta T = \frac{T_{\rm{sys}}\cdot k_{\rm{B}}\cdot \Delta \nu_{\rm{ch}}}{\sqrt{\Delta \nu_{\rm{ch}} \tau}}\;.
\end{equation}

The injected power $P_{\rm cw}$ must exceed this by a factor of at
least the required signal-to-noise ratio of the gain measurement. To
reduce the $1/f$ knee frequency by a factor $F$ requires a correction
of the gain fluctuations to one part in $F/\alpha$ i.e.
\begin{equation}
P_{\rm cw} > \frac{F}{\alpha} k_{\rm B} T_{\rm sys} \sqrt{\frac{\Delta \nu_{\rm ch}}{\tau}}.
\end{equation}

The longest allowed integration time $\tau$ is the inverse of the
unstabilized knee frequency, but shorter integration times are
preferable in order to sample well in to the white-noise dominated
regime. For our receiver with $T_{\rm{sys}}=22\,K$, $\Delta \nu =
500\,\rm{MHz}$ and $\Delta \nu_{\rm{ch}} = 250\,\rm{kHz}$, we use a
sampling time of $\tau_{\rm{ch}} = \frac{1}{23\,\rm{Hz}} =
0.043\,$s. To reduce the $1/f$ knee frequency from the measured value
of 2 Hz to 0.01 Hz we would need a reference CW signal of $P_{\rm{cw}}
> -128$\,dBm. This compares with the total noise power in the full 500
MHz bandwidth of $-98$\, dBm. The CW signal is therefore well below
the noise level in the RF sampled data, and does not contribute to the
dynamic range requirements of the ADC. Although the voltage due to the CW signal can be significantly below the ADC resolution, the effect of the noise-like sky signal overlaid with the CW signal leads to an reduction of the quantization noise of the reference signal and a better calibration \citep{lyons2005reducing}.

Since the gain measurement is not limited by the thermal
noise on the measurement, in practice the uncertainty is dominated by
the fluctuations of the CW signal over the integration time
$\tau$,
\begin{equation}
\frac{\Delta G}{G} = \frac{\Delta P_{\rm{cw}}}{P_{\rm{cw}}}.
\end{equation}

In this case the condition that the effect of residual gain fluctuations is less than the thermal noise is

\begin{equation}
\left(\frac{\Delta P_{\rm{cw}}}{P_{\rm{cw}}}\right)^2 \ll \frac{1}{\tau \cdot \Delta \nu}  \;,
\end{equation}

\noindent and therefore the stability of the cw-signal over the integration time $\tau$ must be

\begin{equation}
\frac{\Delta P_{\rm{cw}}}{P_{\rm{cw}}}\ll \sqrt{\frac{1}{\tau\cdot \Delta \nu}}\;.
\end{equation}

For our particular receiver, and a target corrected knee frequency of
0.01 Hz, this means that we require $\frac{\Delta
  P_{\rm{cw}}}{P_{\rm{cw}}}\ll -54\,\rm{dB}$, or approximately one
part in 200,000. This represents a very high degree of stability,
which can only be achieved with active stabilization of the reference
signal. In practice, provided the stability of the CW signal source is
better than the intrinsic stability of the receiver, the residual
$1/f$ noise spectrum of the receiver data will simply mimic the $1/f$
fluctuation spectrum of the signal source.

\subsection{Laboratory setup}
\label{sec:setup}

The laboratory setup for testing the gain stabilization method is
shown in Figure \ref{fig:fig3}. The input signal for all the
experiments was a temperature stabilized $50\;\Omega$ cold load at
$15$\;K. Following the load is a coupler where the continuous wave
signal is injected into the signal chain. The only other component in
the cold section of the cryostat is a $4$--$8.5$\;GHz low noise
amplifier. Further amplification, filtering and slope compensation is
located outside the cryostat at room temperature. Finally the signal
is down-converted, amplified, filtered, and fed into an
analogue-digital-converter before it is analysed by a digital signal
processing unit. The CW calibration signal is produced by an Anritsu
MG3692B signal generator.

For the digital readout system, we use a ROACH FPGA-based hardware
platform with an iADC digitizer board, developed by the CASPER
collaboration \citep{casper}. The ADCs allow for a maximum sample
frequency of $1\,\rm{GHz}$ with an $8\,\rm{bit}$ resolution over the
voltage range of $\pm 250\,\rm{mV}$.

\subsection{Correlation of gain fluctuations across the frequency band of the receiver}
\label{sec:correlation}

\begin{figure}
 \includegraphics[width=\columnwidth]{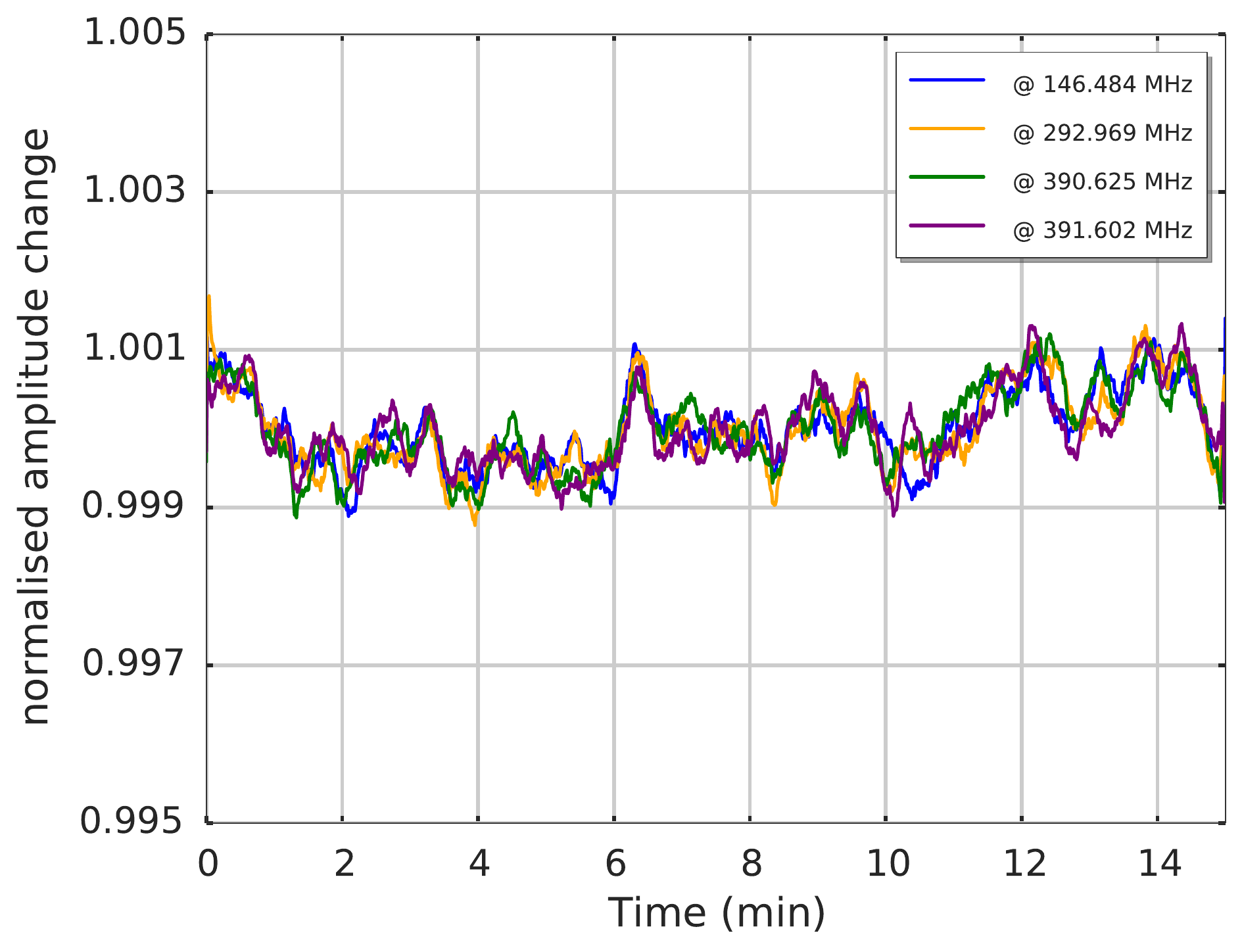}
 \caption{Measured fractional change in signal amplitude of independent frequency channels over $15$\;min, measuring a stabilized cold load.}
 \label{fig:fig4}
\end{figure}

As mentioned in Section \ref{sec:method_overview}, the method presented
here depends on the assumption that the gain fluctuations are
correlated across the frequency band of the receiver. Figure
\ref{fig:fig4} shows the time variations of several different channels
of the frequency spectrum over $15$\;min. Three of the channels shown
are widely separated by approximately $100$\;MHz, while the fourth
channel is only spaced less than $1$\;MHz from the third channel. One
can clearly see the correlation of the variations between different
frequencies and also that the correlation between two widely separated
channels is similar to that between two adjacent channels.

To confirm this estimate we calculated the Pearson correlation
coefficient (PCC) \citep{pcc2009} of the time-ordered frequency data
to see how the gain drift correlates between different frequency
channels. The PCC provides an indication of the correlation between
two variables and is normalised to be within $1$ and $0$, where a
value of $0$ describes an uncorrelated relationship and a value of $1$
describes a perfect correlation.

\begin{figure}
 \includegraphics[width=\columnwidth]{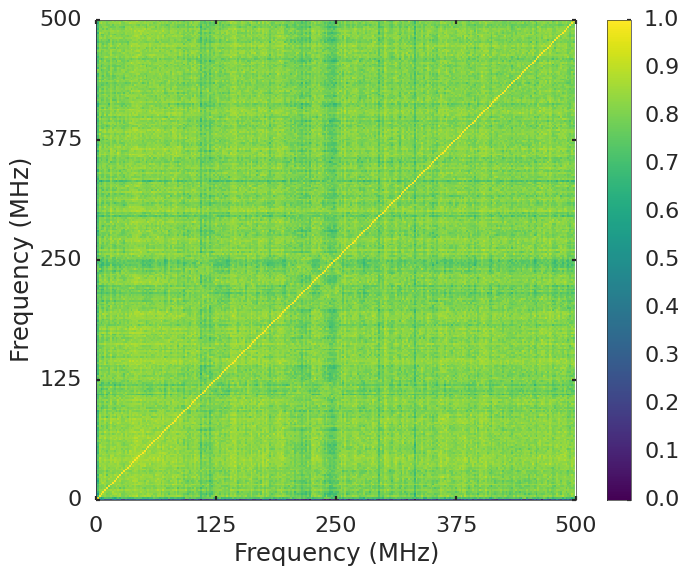}
 \caption{Pearson correlation of the frequency dependent gain drift.}
 \label{fig:fig5}
\end{figure}

The result is shown in Figure \ref{fig:fig5}. The yellow diagonal line
represents the Pearson coefficient for the autocorrelations of the
frequency channels, hence the PCC is $1$. The plot shows no frequency
dependence of the gain drift, which can be seen by the uniform
coefficient values for the different frequency channel
combinations, and a high value of the correlation coefficient for
different channels ($0.7$--$0.9$). The darker lines visible in the
plot indicate channels that have a higher noise level compared to
others, hence the correlation for these channels is lower, but
nevertheless independent of frequency.

We conducted Monte Carlo simulations using LabVIEW to evaluate this result
theoretically. The goal was to calculate the expected correlation
coefficient for two frequency channels whose gain fluctuations are
fully correlated but that are overlaid by uncorrelated white noise.

The simulation setup mirrors the experiment with which the
correlation coefficients in Figure \ref{fig:fig5} were
calculated. Here the bandwidth of $500$\;MHz is split into $512$
channels. The receiver output was recorded for $15$ Minutes and the
data is smoothed to an integration time of $\tau=20,84$\;s, as shown
in Figure \ref{fig:fig4}. The normalised rms of the uncorrelated white
noise on the signal therefore is $\frac{1}{\sqrt{20,84\,\rm{s}\cdot
    0.977\,\rm{MHz}}}=0.00022$.

To simulate the correlation coefficient of two frequency channels we
produced two different Gaussian white noise signals and added a fully
correlated random $1/f$ noise signal to both channels. The normalised
rms of one channel matched the measured average rms of all channels of
the recorded receiver data, here $0.00051$. The simulation was then
repeated several thousand times.

The simulations resulted in an expected correlation coefficient for
our experiment of $r=0.80\pm 0.06$ whereas the average measured
correlation coefficient of the $512\cdot 512$ Matrix is $r=0.81\pm
0.04$ (without considering the values for the autocorrelations in the
matrix).  
This result shows confidently that our assumption is valid that the gain fluctuations of the receiver are fully correlated across
the spectrum.
A control simulation using uncorrelated $1/f$ noise in the two channels resulted in a correlation coefficient value of zero, as one would expect.

\subsection{Laboratory measurements and verification}
\label{sec:lab_measurement}

To find out how efficiently gain fluctuations can be corrected for, we conducted several measurements with the laboratory test setup of Section \ref{sec:setup}. A special firmware implementation for the digital readout was used in order to measure the overall gain drift as well as the gain drift in the single frequency channel containing the CW signal.

The performance of the stabilization method is determined by a comparison between the uncorrected and the corrected power spectrum for a range of different CW signal frequencies and power levels. This is done by using the noise model presented in Section \ref{sec:gain_fluctuations}, which allows us to fit the power spectra and compare the values of the pink noise level. The stability of the CW reference signal over the integration time of the measurements is crucial (see Section \ref{sec:method_overview}), and dominates the residual pink noise of the stabilized receiver system. Therefore the internal active level control of the reference signal generator was used while keeping the device at a very stable temperature using the air conditioning system of the laboratory.

\begin{figure}
 \includegraphics[width=\columnwidth]{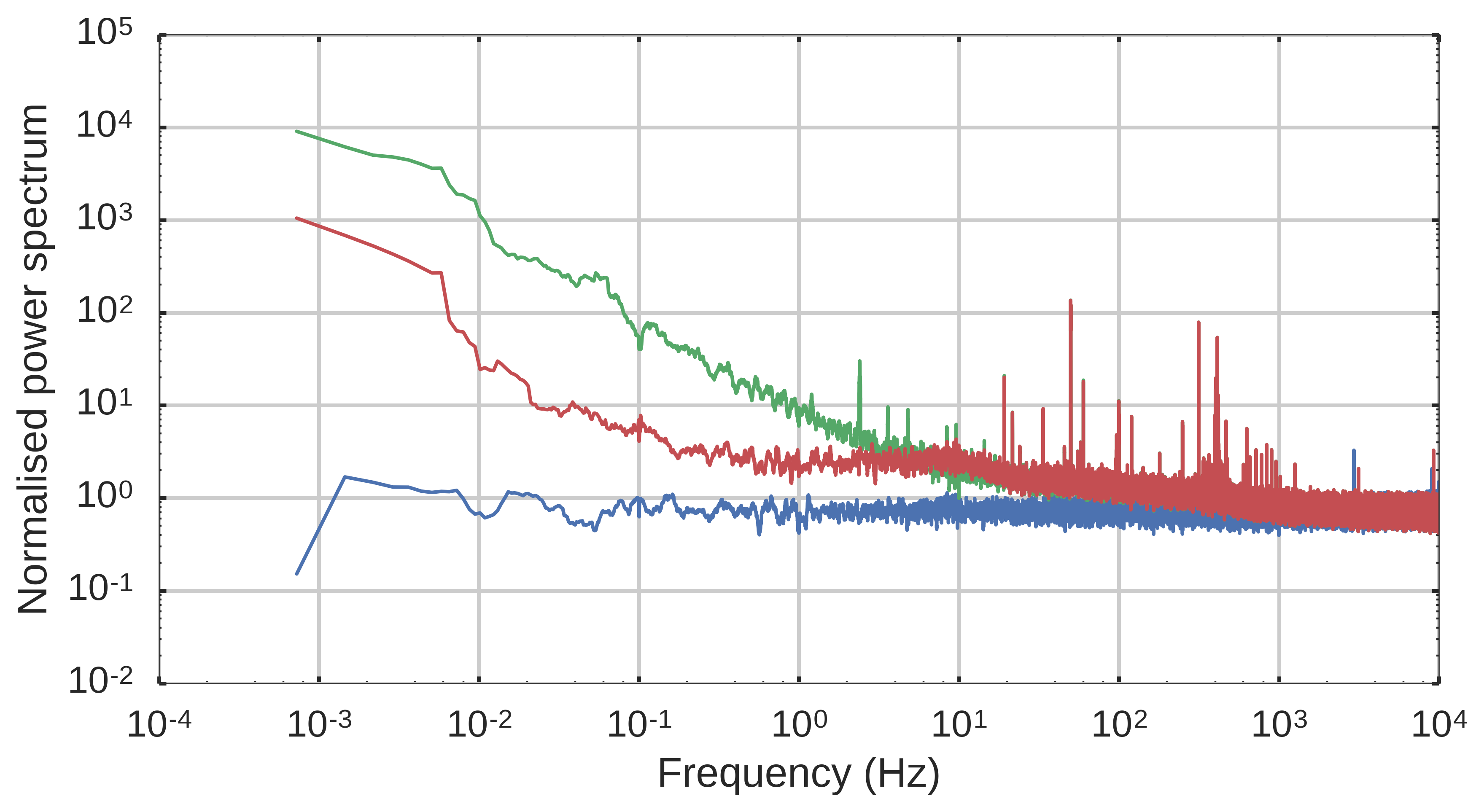}
 \caption{Unstabilised spectrum (green) overlaid with a stabilized spectrum (red) using the cw-signal receiver stabilization method. Also shown is the spectrum obtained by reversing the sign of alternate data samples (blue), which cancels out all correlated signals and shows the true white noise level.  The Figure shows the normalised receiver output spectra for a $50\;\Omega$ cold load as an input, temperature stabilized at $15$\;K. The correction applies for frequencies lower than $11.5$\;Hz. The spectra have been logarithmically smoothed using a moving average filter.
The Figure shows the result of using a $5.125$\;GHz reference signal with a power level of $-21.8$\;dBm.
}
 \label{fig:fig6}
\end{figure}

\begin{figure}
 \includegraphics[width=\columnwidth]{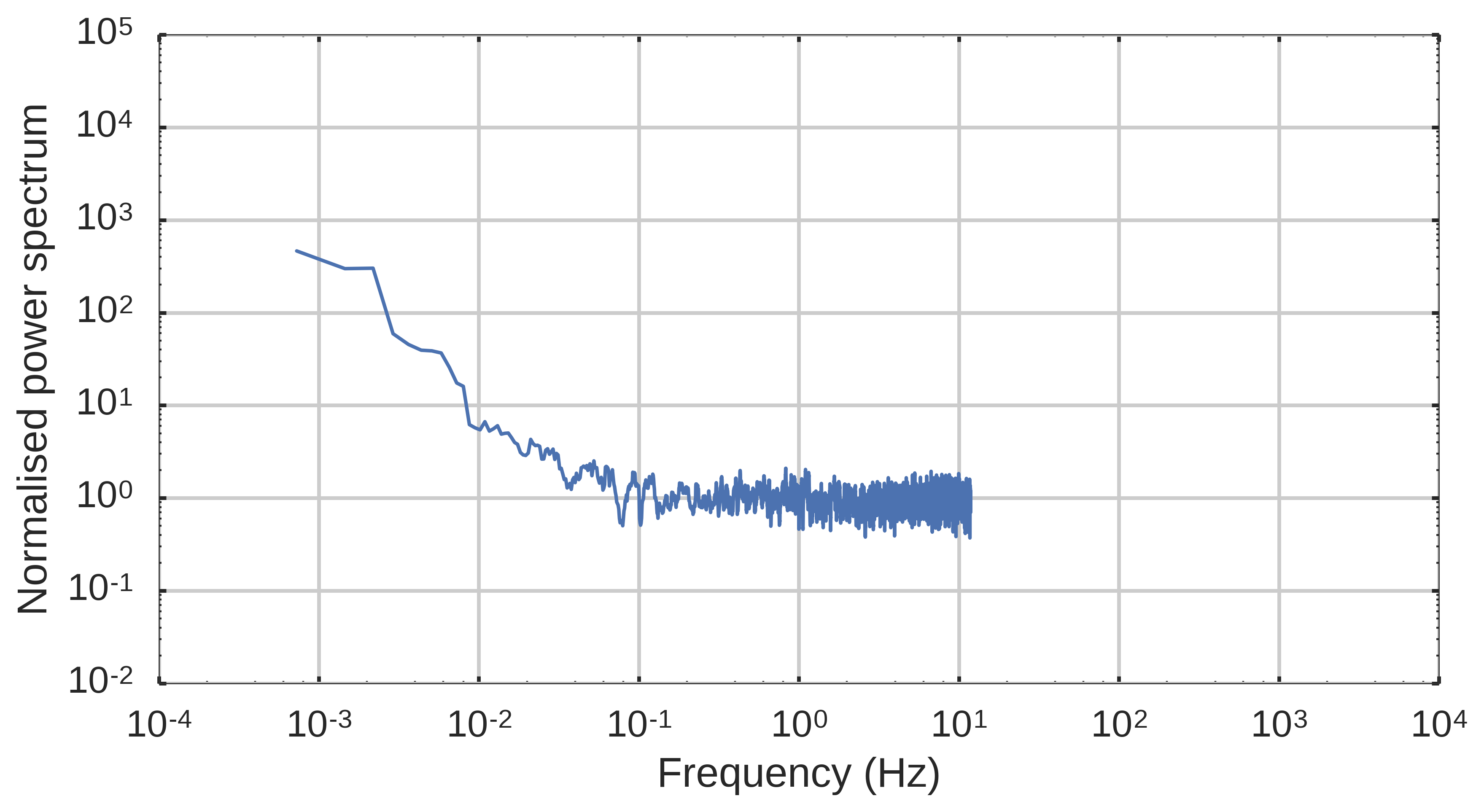}
 \caption{Power spectrum of the fractional power fluctuations (i.e. $\frac{P}{<P>}-1$) of the of the Anritsu MG3692B signal generator measured directly (not through the receiver system) in an independent measurement. The spectrum has been logarithmically smoothed using a moving average filter. 
}
 \label{fig:fig7}
\end{figure}

In Figure \ref{fig:fig6} we show typical examples of corrected and uncorrected noise power spectra. The corresponding fitted noise model values are shown in Table \ref{tab:results}.
We also show the noise power spectrum obtained by reversing the sign of alternate data samples. This corresponds to the ideal case of a very fast Dicke switch with exactly matched load temperature, and as expected shows white noise behaviour down to the lowest frequency.
The measurement in Figure \ref{fig:fig6} was done using cw reference signal at $5.125$\;GHz, which corresponds to an intermediate frequency of $125$\;MHz. The noise power is measured over the full IF bandwidth of 0 -- 500\;MHz. 
The power level of the CW signal was measured with a frequency of $23$\;Hz and therefore pink noise with a frequency below $11.5$\;Hz is corrected for with our experimental setup. As the Figure shows,
the pink noise level at $0.01$\;Hz is reduced by approximately $10$\;dB. 
The white noise level is unaltered except a slight increase around $10$\;Hz due to intrinsic noise on the CW signal. Less stable CW signal generators we used showed an even stronger increase in this frequency range. In addition, at very low frequencies the noise level of the stabilized receiver signal is dominated by the residual $1/f$ noise from the signal generator, shown in Figure \ref{fig:fig7}. The corresponding fitted noise model values for the signal generator are shown in Table \ref{tab:fitted_noise_model_anritsu}. This shows the power spectrum of the fractional power fluctuations of the signal generator measured directly using the same digitization setup but without passing through the receiver system. The power spectrum at low frequencies is at a similar level to the residual gain fluctuations in Figure \ref{fig:fig6}. This emphasizes the need for improved active level control for very long integration times.  Nevertheless, the receiver stability for all frequencies below a few Hertz is substantially increased. We have begun development of an ALC-stabilized CW source for use with the Goonhilly receiver, using careful thermal control of a diode power detector and voltage-controlled attenuator, that should provide better power stability than the bench signal generators used here.


\begin{table}
\centering
\caption{Comparison of the fitted power spectra models, before and after stabilization. Both spectra are both normalized to their white noise level.}
\label{tab:results}
\begin{tabular}{@{}llll@{}}
                                      
\multicolumn{2}{l}{}            & \multicolumn{2}{c}{}                                              \\ \\
 \multicolumn{4}{c}{CW Power -21.8\,dBm, CW Frequency 5.125\,GHz:}                                                                 \\ \hline 
\multicolumn{2}{l}{Parameter}   & \multicolumn{1}{l}{Unstabilized} & \multicolumn{1}{l}{Stabilized}  \\ \hline
\multicolumn{1}{l}{$\alpha$}  &\multicolumn{1}{l|}{}  & \multicolumn{1}{l|}{0.9}             & \multicolumn{1}{l}{1.1} \\
\multicolumn{1}{l}{$\sigma_{w}$}  &\multicolumn{1}{l|}{}  & \multicolumn{1}{l|}{1.0}  & \multicolumn{1}{l}{1.0}         \\
\multicolumn{1}{l}{$\sigma_{c}$}  &\multicolumn{1}{l|}{}  & \multicolumn{1}{l|}{3.0}   & \multicolumn{1}{l}{0.6}    \\
\multicolumn{1}{l}{$P_{xx}$ at $f>100\,\rm{Hz}$}  &\multicolumn{1}{l|}{(dB)}  & \multicolumn{1}{l|}{0.0}   & \multicolumn{1}{l}{0.0}  \\
\multicolumn{1}{l}{$P_{xx}$ at $f=10^{-2}\,\rm{Hz}$} &\multicolumn{1}{l|}{(dB)}  & \multicolumn{1}{l|}{27.5}    & \multicolumn{1}{l}{17.6}  \\


\end{tabular}
\end{table}

\begin{table}
\centering
\caption{Fitted noise model parameters for the spectrum of the Anritsu MG3692B signal generator, shown in Figure \ref{fig:fig7}.}
\label{tab:fitted_noise_model_anritsu}
\begin{tabular}{@{}lll@{}}
\hline                                        
\multicolumn{2}{l}{Parameter}   & \multicolumn{1}{l}{Value}  \\ \hline
\multicolumn{1}{l}{$\alpha$}  &\multicolumn{1}{l|}{}  & \multicolumn{1}{l}{1.7}              \\
\multicolumn{1}{l}{$\sigma_{w}$}  &\multicolumn{1}{l|}{}  & \multicolumn{1}{l}{$1.0$}          \\
\multicolumn{1}{l}{$\sigma_{c}$}  &\multicolumn{1}{l|}{}  & \multicolumn{1}{l}{$0.07$}      \\
\multicolumn{1}{l}{$P_{xx}$ at $f>1\,\rm{Hz}$}  &\multicolumn{1}{l|}{(dB)}  & \multicolumn{1}{l}{0.0}  \\
\multicolumn{1}{l}{$P_{xx}$ at $f=10^{-2}\,\rm{Hz}$} &\multicolumn{1}{l|}{(dB)}  & \multicolumn{1}{l}{10.9}  \\

\end{tabular}
\end{table}

\section{Conclusions}
\label{sec:conclusion}

In this paper we present a gain stabilization method for a radio astronomy receiver using a continuous-wave calibration signal. It is especially suitable for receivers with a digital back-end with high spectral resolution, and does not require especially high dynamic range in the ADC. A narrow band amplitude-stabilized CW signal is coupled into the signal chain before the first low noise amplifier and its amplitude is tracked in a single spectral channel. The variations of this channel's power level are used to correct the observed signal for fluctuations in the overall gain of the receiver.

In the main section of this paper we presented a discussion of the origin and types of receiver noise, and a noise model to characterize receiver behaviour. The stabilization method was tested and verified using a newly developed $4$--$8.5$\;GHz heterodyne receiver with a digital back-end in a laboratory setup. The results confirm the effectiveness of the stabilization. The $1/f$ noise level at $0.01$\;Hz is reduced by approximately 10\;dB. The results depend on the fact that gain fluctuations are strongly correlated across the full frequency band which we can show by calculating the Pearson correlation of the fluctuations between all the frequency channels. Their values closely match simulation results that assume fully correlated frequency channels overlaid with uncorrelated white noise.

Whereas many stabilization methods come with the cost of reduced receiver sensitivity, usually by a factor of $\sqrt{2}$ to $2$ due to the use of a noise-like reference signal, this method only loses the sky signal in a single frequency channel. Therefore receiver sensitivity is essentially unaltered for a system with high spectral resolution. This method requires a stabilized CW signal injected into the signal chain and a high channel isolation in the computation of the resulting spectrum. Existing systems which use a broadband noise signal for stabilization or calibration, and hence have a suitable signal injection point early in the RF chain, could readily be modified to use this method. 

\section*{Acknowledgements}

The stabilization method presented in this paper was developed for the Consortium of Universities for Goonhilly Astronomy (CUGA) and for the Goonhilly Earth Station (GES), supported by an STFC CASE studentship. We would also like to thank Luke Jew for use of his noise model code and useful discussions about power spectra.




\bibliographystyle{mnras}
\bibliography{mnras_gain_stabilisation}

\begin{thebibliography}{}
\makeatletter
\relax
\def\mn@urlcharsother{\let\do\@makeother \do\$\do\&\do\#\do\^\do\_\do\%\do\~}
\def\mn@doi{\begingroup\mn@urlcharsother \@ifnextchar [ {\mn@doi@}
  {\mn@doi@[]}}
\def\mn@doi@[#1]#2{\def\@tempa{#1}\ifx\@tempa\@empty \href
  {http://dx.doi.org/#2} {doi:#2}\else \href {http://dx.doi.org/#2} {#1}\fi
  \endgroup}
\def\mn@eprint#1#2{\mn@eprint@#1:#2::\@nil}
\def\mn@eprint@arXiv#1{\href {http://arxiv.org/abs/#1} {{\tt arXiv:#1}}}
\def\mn@eprint@dblp#1{\href {http://dblp.uni-trier.de/rec/bibtex/#1.xml}
  {dblp:#1}}
\def\mn@eprint@#1:#2:#3:#4\@nil{\def\@tempa {#1}\def\@tempb {#2}\def\@tempc
  {#3}\ifx \@tempc \@empty \let \@tempc \@tempb \let \@tempb \@tempa \fi \ifx
  \@tempb \@empty \def\@tempb {arXiv}\fi \@ifundefined
  {mn@eprint@\@tempb}{\@tempb:\@tempc}{\expandafter \expandafter \csname
  mn@eprint@\@tempb\endcsname \expandafter{\@tempc}}}

\bibitem[\protect\citeauthoryear{Benesty, Chen, Huang  \& Cohen}{Benesty
  et~al.}{2009}]{pcc2009}
Benesty J.,  Chen J.,  Huang Y.,   Cohen I.,  2009, Noise Reduction in Speech
  Processing.
Springer, Berlin Heidelberg

\bibitem[\protect\citeauthoryear{{Blum}}{{Blum}}{1959}]{1959AnAp...22..140B}
{Blum} E.~J.,  1959, Annales d'Astrophysique, \href
  {http://adsabs.harvard.edu/abs/1959AnAp...22..140B} {22, 140}

\bibitem[\protect\citeauthoryear{Colvin}{Colvin}{1961}]{colvin1961}
Colvin R.~S.,  1961, A Study of Radio Astronomy Receivers, Stanford Radio
  Astronomy Institute Publication No. 18A.
Stanford Electronics Labs., Stanford University

\bibitem[\protect\citeauthoryear{Crochiere \& Rabiner}{Crochiere \&
  Rabiner}{1983}]{crochiere1983multirate}
Crochiere R.,  Rabiner L.,  1983, Multirate Digital Signal Processing.
Prentice-Hall

\bibitem[\protect\citeauthoryear{Dicke}{Dicke}{1946}]{dicke1946}
Dicke R.~H.,  1946, Review of Scientific Instruments, 17, 268

\bibitem[\protect\citeauthoryear{Heywood et~al.,}{Heywood
  et~al.}{2011}]{heywood2011expanding}
Heywood I.,  et~al., 2011, arXiv preprint arXiv:1103.1214

\bibitem[\protect\citeauthoryear{{Jarosik} et~al.,}{{Jarosik}
  et~al.}{2003}]{2003ApJS..145..413J}
{Jarosik} N.,  et~al., 2003, \mn@doi [\apjs] {10.1086/346080}, \href
  {http://adsabs.harvard.edu/abs/2003ApJS..145..413J} {145, 413}

\bibitem[\protect\citeauthoryear{Jew}{Jew}{2017}]{jew2017}
Jew L.,  2017, PhD thesis, University of Oxford

\bibitem[\protect\citeauthoryear{{Jones} et~al.,}{{Jones}
  et~al.}{2018}]{projectpaper}
{Jones} M.~E.,  et~al., 2018, \mn@doi [\mnras] {10.1093/mnras/sty1956}, \href
  {http://adsabs.harvard.edu/abs/2018MNRAS.480.3224J} {480, 3224}

\bibitem[\protect\citeauthoryear{Kelly, Lyons  \& Root}{Kelly
  et~al.}{1963}]{10.2307/2098930}
Kelly E.~J.,  Lyons D.~H.,   Root W.~L.,  1963, Journal of the Society for
  Industrial and Applied Mathematics, 11, 235

\bibitem[\protect\citeauthoryear{King et~al.,}{King et~al.}{2013}]{king2013c}
King O.,  et~al., 2013, Monthly Notices of the Royal Astronomical Society, 438,
  2426

\bibitem[\protect\citeauthoryear{Lyons \& Yates}{Lyons \&
  Yates}{2005}]{lyons2005reducing}
Lyons R.,  Yates R.,  2005, Microwaves RF, 44, 72

\bibitem[\protect\citeauthoryear{Mennella et~al.,}{Mennella
  et~al.}{2003}]{mennella2003advanced}
Mennella A.,  et~al., 2003, arXiv preprint astro-ph/0307116

\bibitem[\protect\citeauthoryear{Naldi et~al.,}{Naldi et~al.}{2017}]{TPM}
Naldi G.,  et~al., 2017, Journal of Astronomical Instrumentation, 6, 1641014

\bibitem[\protect\citeauthoryear{Pollak}{Pollak}{2018}]{pollak2018}
Pollak A.~W.,  2018, PhD thesis, University of Oxford

\bibitem[\protect\citeauthoryear{{Press}}{{Press}}{1978}]{1978ComAp...7..103P}
{Press} W.~H.,  1978, Comments on Astrophysics, \href
  {http://adsabs.harvard.edu/abs/1978ComAp...7..103P} {7, 103}

\bibitem[\protect\citeauthoryear{Price}{Price}{2016}]{price2016}
Price D.~C.,  2016, ArXive e-prints: 1607.03579, \href
  {https://arxiv.org/abs/1607.03579} {}

\bibitem[\protect\citeauthoryear{{Santos} et~al.,}{{Santos}
  et~al.}{2015}]{2015aska.confE..19S}
{Santos} M.,  et~al., 2015, Advancing Astrophysics with the Square Kilometre
  Array (AASKA14), \href {http://adsabs.harvard.edu/abs/2015aska.confE..19S}
  {p.~19}

\bibitem[\protect\citeauthoryear{Schafer \& Rabiner}{Schafer \&
  Rabiner}{1973}]{schafer1973design}
Schafer R.,  Rabiner L.,  1973, IEEE Transactions on Audio and
  Electroacoustics, 21, 165

\bibitem[\protect\citeauthoryear{{Seiffert}, {Mennella}, {Burigana},
  {Mandolesi}, {Bersanelli}, {Meinhold}  \& {Lubin}}{{Seiffert}
  et~al.}{2002}]{2002A&A...391.1185S}
{Seiffert} M.,  {Mennella} A.,  {Burigana} C.,  {Mandolesi} N.,  {Bersanelli}
  M.,  {Meinhold} P.,   {Lubin} P.,  2002, \mn@doi [\aap]
  {10.1051/0004-6361:20020880}, \href
  {http://adsabs.harvard.edu/abs/2002A%26A...391.1185S} {391, 1185}

\bibitem[\protect\citeauthoryear{Wait}{Wait}{1967}]{wait1967}
Wait D.~F.,  1967, Journal of Research of the Notional Bureau of Standards --
  C. Engineering and Instrumentation, 71C

\bibitem[\protect\citeauthoryear{Werthimer}{Werthimer}{2011}]{casper}
Werthimer D.,  2011, in 2011 XXXth URSI general assembly and scientific
  symposium. pp~1--4

\makeatother
\end{thebibliography}




%
%


\bsp	
\label{lastpage}
\end{document}